\documentclass[
 aps,pre,reprint,superscriptaddress,
 amsmath,amssymb,longbibliography
]{revtex4-2}
\usepackage{graphicx}
\usepackage{pgf}

\newsavebox{\resultfigurebox}

\usepackage{bm}
\usepackage{hyperref}
\hypersetup{hidelinks}
\newcommand{\ii}{\mathrm{i}}
\newcommand{\Rea}{\operatorname{Re}}
\newcommand{\Ima}{\operatorname{Im}}
\newcommand{\conj}[1]{\overline{#1}}
\begin{document}

\title{Noise-driven loss of chimera states and bifurcations in a nonlocal oscillator ring}
\author{Lintao Liu}
\author{Nariya Uchida}%
\email{uchida@cmpt.phys.tohoku.ac.jp}
\affiliation{Department of Physics, Tohoku University, Sendai 980-8578, Japan}
\date{\today}
\begin{abstract}
Independent phase noise can remove the spatial contrast of a chimera while leaving substantial collective coherence. We study this transition in a nonlocally coupled Kuramoto--Sakaguchi ring with a cosine kernel using the nonlinear Fokker--Planck equation. Stationary self-consistency and spatial spectra identify two bifurcations of the uniform ordered state at a reference phase lag. A symmetry-aware amplitude reduction distinguishes a subcritical lower endpoint $D_l$ from a supercritical upper endpoint $D_u$ and shows that reconstruction of the collective rotation frequency is essential to obtain the cubic coefficient. The upper endpoint admits a nonequilibrium Landau-type description of continuous spatial translation-symmetry restoration. Direct integration confirms stationary modulation below the upper threshold, algebraic relaxation at criticality, and exponential decay toward uniform partial synchronization above it. Continuation in noise intensity and phase lag resolves finite-amplitude saddle nodes, a narrow cusp, and a distinct cubic-zero point with a positive quintic coefficient. Together, these results connect the bifurcation structure of noisy chimera branches to spatial symmetry restoration and critical relaxation.
\end{abstract}
\maketitle

\section{Introduction}\label{sec:introduction}
Chimera states \cite{kuramoto2002coexistence,PhysRevLett.93.174102,Panaggio_2015} combine spatial regions of different coherence in networks of identical oscillators. In nonlocally coupled phase-oscillator rings, their continuum description is a rotating statistical profile determined by a self-consistent coupling field \cite{Abrams2006, Omelchenko2018}. Independent phase noise broadens the local phase distributions and changes the stability of the collective pattern. The resulting loss of spatial contrast need not coincide with the loss of global phase coherence. Understanding this distinction requires following both stationary densities and their perturbations.

Several ingredients of this problem have established precedents. Laing studied chimera bifurcations in heterogeneous oscillator networks, including a cosine-kernel ring with distributed natural frequencies \cite{Laing2009,LAING20091569}. Independent Gaussian phase noise and full Fourier representations of coupled Fokker--Planck equations were subsequently considered in two-population chimera models \cite{Laing2012}. 
Circular-cumulant reductions describe the effect of weak Gaussian noise beyond the Ott--Antonsen manifold \cite{Tyulkina2018}, and bifurcation calculations for noisy two-population chimeras include the Gaussian case \cite{Dolmatova2023}. 
Masoliver et al. examined chimera dynamics in a nonlocally coupled oscillator ring, including robustness to independent noise in the context of phase precession \cite{10.3389/fncir.2025.1634298}.
These works show that noisy chimera dynamics and noise-dependent bifurcations are established subjects. Gaussian phase diffusion nevertheless changes the full moment hierarchy through damping proportional to the square of the phase harmonic index, so a bifurcation diagram obtained for frequency heterogeneity cannot be transferred directly to the present model.

The spatial ring adds a continuous translation symmetry and a distinction between uniform and spatially modulated densities. Noisy nonlocal oscillator systems can exhibit spatial instabilities and time-dependent collective behavior \cite{Kawamura}. A quantitative account of noise-driven loss of spatial structure must therefore distinguish a linear instability of the uniform state from disappearance of a finite-amplitude stationary branch.

Here we determine the two spatial instability thresholds of a uniform ordered density, calculate the nonlinear coefficients that distinguish their local bifurcations, and test the predicted relaxation directly in the kinetic equation. Continuation in noise intensity and phase lag then resolves how continuous loss of modulation connects to finite-amplitude folds and a narrow region of multistability. The focus is the quantitative organization of these phenomena for independent Gaussian phase noise on the specified cosine-kernel branch family.

We call states continued from the classical chimera family noisy chimera states when their local coherence and phase-slip current exhibit the corresponding spatial contrast. At nonzero diffusion, coherence is statistical rather than perfect phase locking. We use nonuniform for general rotating densities and weakly modulated branches near spatial bifurcations; modulation alone is not a chimera criterion. Uniform ordered states are spatially uniform, partially synchronized densities, distinct from the phase-incoherent density.

Section~\ref{sec:model} introduces the model and observables; the appendices collect detailed methods and convergence checks. Section~\ref{sec:stationary} constructs rotating states and identifies their spatial stability thresholds. Section~\ref{sec:amplitude} derives and validates the local amplitude description. Section~\ref{sec:continuation} follows its extension into the two-parameter diagram. Section~\ref{sec:conclusions} discusses the physical implications and scope.

\section{Model and kinetic description}\label{sec:model}
\subsection{Microscopic dynamics and mean-field limit}
Consider fixed positions $x_i=i/N$ on the unit ring and the Kuramoto--Sakaguchi equations
\begin{equation}\label{eq:microscopic}
\begin{aligned}
d\theta_i &= \biggl[ \omega + \frac{K}{N} \sum_{j=1}^N G(x_i - x_j) \\
&\quad \times \sin(\theta_j - \theta_i - \alpha) \biggr] dt + \sqrt{2D} \, dW_i
\end{aligned}
\end{equation}
Here $K>0$, the natural frequency $\omega$ is common to all oscillators, and the Wiener processes are independent:
$\mathbb E[dW_i\,dW_j]=\delta_{ij}\,dt$. The coefficient $D$ is the phase diffusivity. We use the normalized cosine kernel
\begin{equation}\label{eq:kernel}
 \begin{gathered}
 G(x)=1+A\cos(2\pi x),\qquad 0\le A\le1,\\
 \int_0^1G(x)\,dx=1,\qquad \alpha=\pi/2-\beta.
 \end{gathered}
\end{equation}
The finite-$N$ process has an exact $N$-body Fokker--Planck equation. Its closed, nonlinear one-body description additionally assumes the dense nonlocal mean-field limit: the number of coupled oscillators grows with $N$, while the kernel and its $1/N$ normalization remain fixed. This is the continuum approximation used here.

Let $f(x,\theta,t)$ be the conditional phase density, with $f\ge0$ and $\int_0^{2\pi}f\,d\theta=1$ at each $x$. Define the local moment and the smoothed coupling field by
\begin{align}
 z(x,t)&=\int_0^{2\pi}e^{\ii\theta}f(x,\theta,t)\,d\theta,
 \label{eq:local-moment}\\
 Z(x,t)&=\int_0^1G(x-y)z(y,t)\,dy.
 \label{eq:nonlocal-field}
\end{align}
The local coherence $|z|$ and the nonlocal field $|Z|$ are different observables. The self-consistent angular drift is
\begin{equation}\label{eq:drift}
 v[f](x,\theta,t)=\omega+K\Ima\!\left[e^{-\ii(\theta+\alpha)}Z(x,t)\right].
\end{equation}
In the dense mean-field limit, It\^o calculus gives the Fokker-Planck equation
\begin{equation}\label{eq:fp}
 \boxed{\partial_t f=-\partial_\theta\bigl(v[f]f\bigr)
                    +D\partial_\theta^2 f.}
\end{equation}
Both variables have periodic boundary conditions. Noise diffuses the phase, while the positions $x_i$ remain fixed; there is no spatial diffusion term. Eq.~\eqref{eq:fp} conserves the phase normalization at each position.

\subsection{Rotating-frame Fourier hierarchy}
From here onward we remove the common natural frequency and measure time in units of $K^{-1}$, so that $\omega=0$, $K=1$, and the symbol $D$ denotes the normalized diffusivity $D/K$. We fix $A=0.95$. The reference branch and relaxation calculations use $\beta=0.2$; Sec.~\ref{sec:continuation} varies this parameter.

In a frame rotating at $\Omega$, let $\vartheta=\theta-\Omega t$ and $f_\Omega(x,\vartheta,t)=f(x,\vartheta+\Omega t,t)$. Define
\begin{equation}\label{eq:phase-moments}
 \begin{gathered}
 a_n(x,t)=\int_0^{2\pi}e^{\ii n\vartheta}f_\Omega(x,\vartheta,t)\,d\vartheta,\\
 a_0=1,\qquad a_{-n}=\conj{a_n}.
 \end{gathered}
\end{equation}
We reuse $z=a_1$ and $Z=G*a_1$ for the fields in this frame. Applying the kinetic equation to the phase harmonics gives the exact hierarchy
\begin{equation}\label{eq:hierarchy}
 \begin{aligned}
 \dot a_n={}&(-Dn^2-\ii n\Omega)a_n\\
 &+\frac n2\left(e^{-\ii\alpha}Za_{n-1}
              -e^{\ii\alpha}\conj Z a_{n+1}\right).
 \end{aligned}
\end{equation}
We truncate at $n=M$ with $a_{M+1}=0$ and check both moment convergence and reconstructed-density nonnegativity; details are given in Appendix~\ref{sec:kinetic-details}.

The standard Ott--Antonsen ansatz $a_n=z^n$ \cite{OttAntonsen2008} is generally not invariant under Gaussian phase diffusion. A first-moment noise term $\dot z=-Dz$ would generate $-nDz^n$ under that ansatz, whereas Eq.~\eqref{eq:hierarchy} contains $-n^2Da_n$. We retain the full hierarchy rather than imposing this closure. Circular-cumulant expansions provide a systematic weak-noise alternative \cite{Tyulkina2018}, but are not used to obtain the numerical branches considered here.

Writing $\langle h\rangle_x=\int_0^1h(x)\,dx$, the cosine kernel reduces the spatial convolution exactly to
\begin{equation}\label{eq:three-spatial-moments}
 \begin{aligned}
 Z(x,t)={}&c(t)+b_{\cos}(t)\cos(2\pi x)\\
         &+b_{\sin}(t)\sin(2\pi x),\\
 c={}&\langle a_1\rangle_x,\\
 b_{\cos}={}&A\langle a_1\cos(2\pi x)\rangle_x,\\
 b_{\sin}={}&A\langle a_1\sin(2\pi x)\rangle_x.
 \end{aligned}
\end{equation}
This reduction concerns the coupling field only; the density and its phase moments can retain higher spatial harmonics. 

\begin{figure*}[t]
\centering
  \begingroup
  \begin{lrbox}{\resultfigurebox}%
    \input{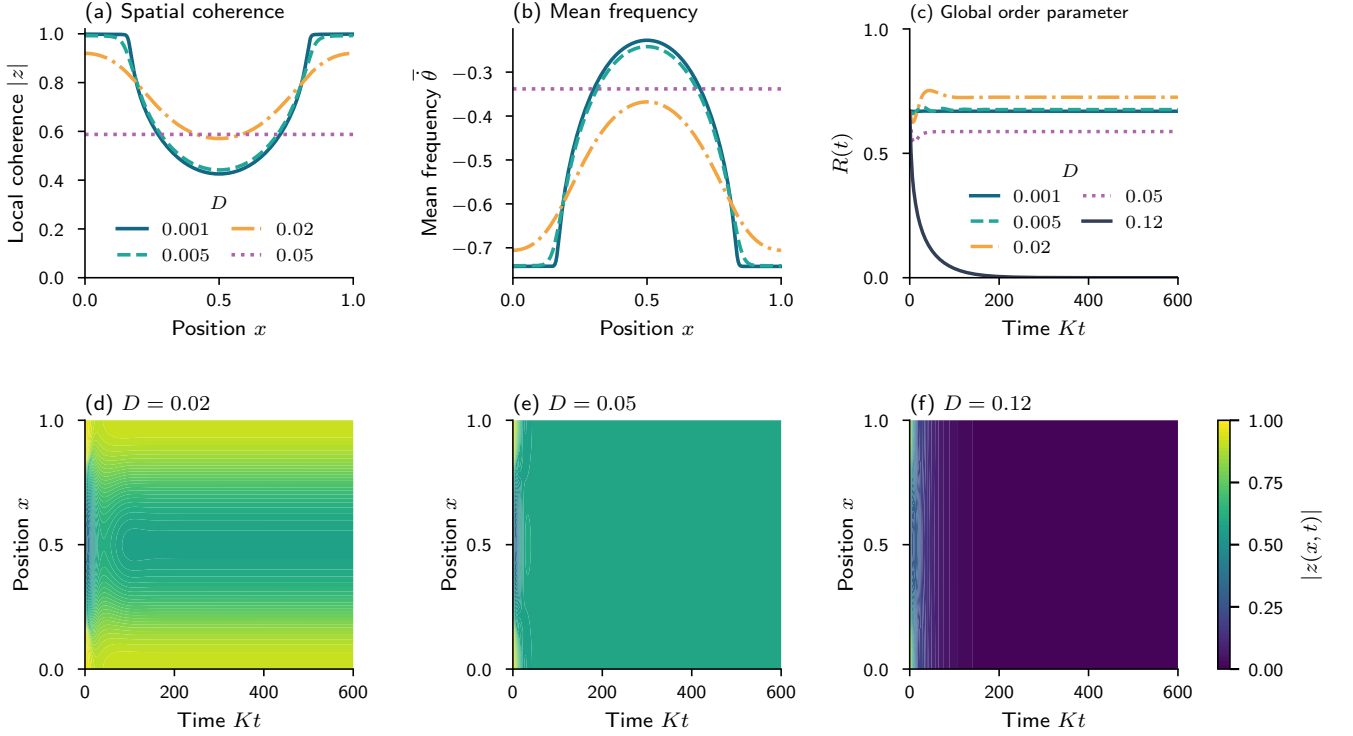}%
  \end{lrbox}%
  \resizebox{\textwidth}{!}{\usebox{\resultfigurebox}}%
  \endgroup

\caption{Spatial structure and representative relaxation at $A=0.95$, $\beta=0.2$, and $K=1$. (a,b) Stationary local coherence and mean angular frequency, with the common natural frequency removed. Matching curves identify $D=0.001$, $0.005$, $0.02$, and $0.05$; the first three are noisy chimera states and the last is a uniform ordered state. (c) Global order parameter $R(t)=|\langle z(x,t)\rangle_x|$ for all five noise intensities, including $D=0.12$. (d)--(f) Direct Fokker--Planck evolution at $D=0.02$, $0.05$, and $0.12$: spatial contrast persists, uniform order develops, and incoherence develops, respectively. Colors show $|z(x,t)|$ on a shared scale from 0 to 1. All trajectories start from the same perturbed noisy chimera at $D_0=0.001$; only $D$ changes. Figure~\ref{fig:profiles-dynamics-supplement} supplies complementary profiles and trajectories; numerical protocols are given in Appendix~\ref{sec:kinetic-details}.}
\label{fig:profiles-dynamics}
\end{figure*}

\subsection{Observables and numerical integration}
We use two complementary scalar observables,
\begin{equation}\label{eq:observables}
 \begin{aligned}
 R(t)&=|c(t)|,\\
 S(t)&=\left(|b_{\cos}(t)|^2+|b_{\sin}(t)|^2\right)^{1/2}.
 \end{aligned}
\end{equation}
Here $R$ measures global phase coherence and $S$ measures modulation of the coupling field; both are invariant under a global phase rotation or spatial translation. For the centered reflection-symmetric stationary family, $S=|b|$. A uniform ordered density has $S=0$ and $R>0$, whereas incoherence has $R=S=0$. The spatial profiles of $|z|$ and the local mean frequency provide additional information not contained in these two scalars. In particle calculations, the corresponding spatial moments are empirical averages over the oscillator positions.

Time evolution uses fourth-order exponential time differencing (ETDRK4) \cite{Kassam2005}; stationary densities and spectra are obtained from the same Fourier hierarchy. Discretization, density checks, and benchmark protocols are collected in Appendix~\ref{sec:kinetic-details}.

Fig.~\ref{fig:profiles-dynamics} illustrates the spatial observables, the global order parameter, and three representative relaxation outcomes. Additional phase densities and trajectories are collected in Fig.~\ref{fig:profiles-dynamics-supplement} in Appendix~\ref{sec:kinetic-details}. The stationary construction, probability current, and stability analysis are developed in Sec.~\ref{sec:stationary}.

\section{Stationary states, spatial stability, and reference branches}\label{sec:stationary}
\subsection{Probability current and local frequency}
A rotating stationary state has $f_\Omega=f_*(x,\vartheta)$ and coupling field $Z_*(x)$. Its drift and phase probability current satisfy
\begin{align}
 v_{\Omega,*}&=-\Omega+\Ima\!\left[e^{-\ii(\vartheta+\alpha)}Z_*(x)\right],
 \label{eq:rotating-drift}\\
 \mathcal J(x)&=v_{\Omega,*}f_*-D\partial_\vartheta f_* ,
 \qquad \partial_\vartheta\mathcal J=0.
 \label{eq:stationary-current}
\end{align}
Thus stationarity makes the current independent of phase, but does not require it to vanish. The mean angular velocity of the unwrapped oscillator phase, in the laboratory frame with its common natural frequency removed, is
\begin{equation}\label{eq:local-frequency}
 \begin{aligned}
 \overline{\dot\theta}(x)
 &=\Omega+2\pi\mathcal J(x)\\
 &=\Ima\!\left[e^{-\ii\alpha}Z_*(x)\conj{a_1(x)}\right].
 \end{aligned}
\end{equation}
Retaining this current includes phase slips and drifting populations. At $D>0$, narrow smooth distributions replace the deterministic locked delta components; local coherence and current describe the resulting statistical contrast across the ring.

\subsection{Local stationary response and field closure}
To construct the reflection-symmetric stationary family, fix its spatial center and global phase and write
\begin{equation}\label{eq:stationary-field}
 Z_*(x)=c+b\cos(2\pi x),\qquad c>0,\quad b\in\mathbb C.
\end{equation}
The local stationary response of Eq.~\eqref{eq:hierarchy} closes the cosine-kernel self-consistency construction \cite{Abrams2006}:

\begin{equation}\label{eq:self-consistency}
 c=\langle a_1\rangle_x,\qquad
 b=A\langle a_1\cos(2\pi x)\rangle_x.
\end{equation}
We solve for $(c,\Rea b,\Ima b,\Omega)$ with the global phase fixed by $c>0$. Reflection symmetry selects the stationary family, while stability and time evolution retain even and odd perturbations. The continued-fraction implementation is given in Appendix~\ref{sec:kinetic-details}.

Let $U(D)=(U_n(D))_{n\ge1}$ denote the spatially uniform ordered rotating density, with frequency $\Omega_h(D)$ and phase condition $U_1=c_h>0$ real. This state has $b=0$ but generally has nonzero higher phase moments. It is distinct from the phase-incoherent density. Linearized operators and adjoint pairings below use the real-vector representation $(\Rea a,\Ima a)$.

\subsection{Uniform-state spectra and spatial thresholds}\label{sec:kinetic-stability}
For the phase-incoherent density $f_{\rm inc}=1/(2\pi)$, the leading growth rate is $-D+\sin\beta/2$ in the parameter range considered. Thus the incoherent stability threshold is

\begin{equation}\label{eq:incoherent-threshold}
 D_{\rm inc}=\tfrac12\sin\beta.
\end{equation}
In dimensional units this is $D_{\rm inc}=K\sin\beta/2$. At $\beta=0.2$, the normalized threshold is $0.0993347$. The spatial thresholds of the ordered state are obtained separately below.

For the uniform ordered density, $L_m$ denotes the linearized block for spatial harmonic $m$. Its real-vector form, including coupling to the conjugate perturbation, is given in Appendix~\ref{sec:kinetic-details}. The neutral global-phase mode belongs to $m=0$ and is separate from the spatial instability.

For each $D$, we first solve the uniform self-consistency problem for $U(D)$ and $\Omega_h(D)$, and then evaluate the leading growth rate of the real-linear $m=1$ block,
\begin{equation}\label{eq:spatial-growth-rate}
 \Lambda_1(D)=\max_{\lambda\in\sigma[L_1(D)]}\Rea\lambda.
\end{equation}
At $\beta=0.2$, its lower negative-to-positive crossing defines $D_l$, and its upper positive-to-negative crossing defines $D_u$. Numerical root finding, checked at $M=64,96,128,192$, gives
\begin{equation}\label{eq:spatial-thresholds}
 \begin{aligned}
 D_l&\simeq0.00446591,\\
 D_u&\simeq0.02649537.
 \end{aligned}
\end{equation}
The critical eigenvalues are real in the rotating frame. The ordered state is unstable to $m=\pm1$ perturbations between these two crossings and stable to these modes immediately outside them. Unlike the incoherent threshold, these values depend on the nonzero higher phase moments of the ordered density and are not obtained from Eq.~\eqref{eq:incoherent-spectrum}. 
Coexisting finite-amplitude states must be followed independently. We use $D_c$ for either spatial endpoint when presenting their common local reduction.

A nonuniform state couples spatial harmonics, so its stability is assessed using the full kinetic operator rather than the uniform-state map. We remove the global-phase and translation neutral modes and check the remaining spectrum. Appendix~\ref{sec:kinetic-details} specifies the operator, constraints, and resolution limits.

\subsection{Phase-diagram overview and relaxation}
\begin{figure*}[t]
\centering
  \begingroup
  \begin{lrbox}{\resultfigurebox}%
    \input{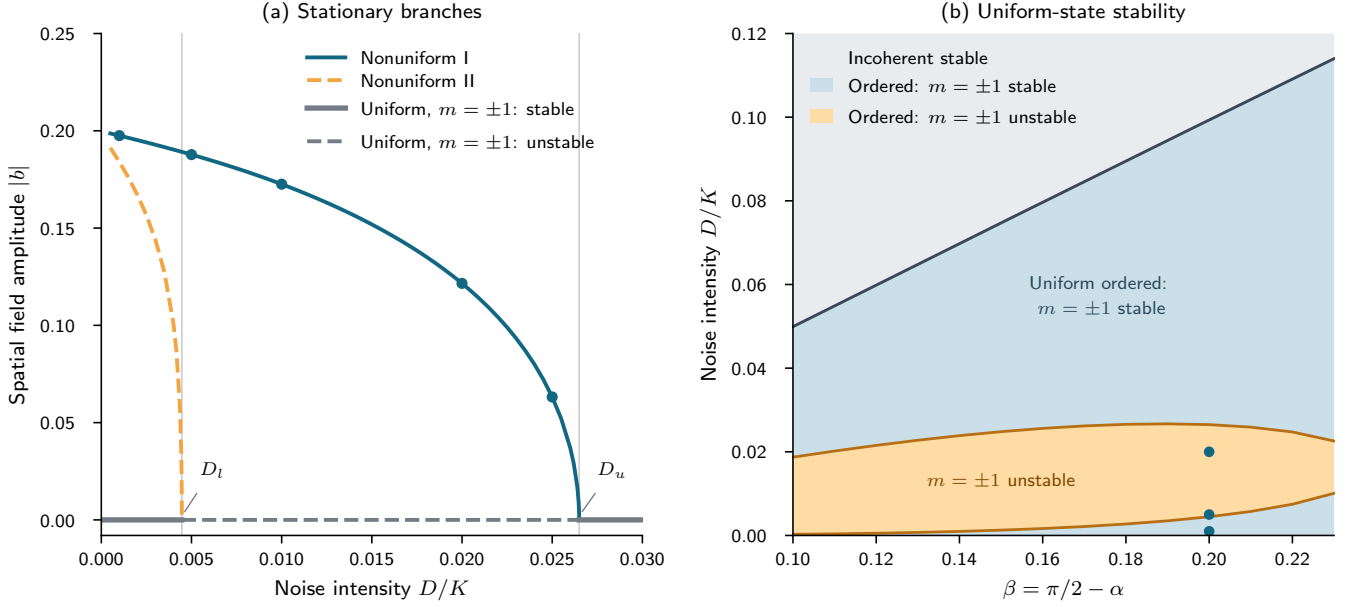}%
  \end{lrbox}%
  \resizebox{\textwidth}{!}{\usebox{\resultfigurebox}}%
  \endgroup

\caption{Stationary branches and uniform-state stability for the cosine kernel with $A=0.95$ and $K=1$. (a) Spatial field amplitude $|b|$ along two continued nonuniform families at $\beta=0.2$. Branch I joins the uniform ordered state at $D_u\simeq0.02649537$, and branch II at $D_l\simeq0.00446591$; colored line styles identify the families, whose stability is assessed separately. On the $|b|=0$ axis, solid and dashed gray segments denote uniform ordered states stable and unstable to $m=\pm1$ perturbations, respectively. (b) Uniform-state stability over $0.10\le\beta\le0.23$: the phase-incoherent state is stable above the exact boundary $D_{\rm inc}=\sin\beta/2$; below this boundary the uniform ordered state is unstable to $m=\pm1$ perturbations in the orange band and stable to these modes in the blue regions. Colored dots mark selected nonuniform solutions at $\beta=0.2$, including one in the region where uniform order is stable to these modes. The map classifies uniform-state linear stability in the indicated modes; it does not classify all coexisting nonuniform states.}
\label{fig:overview}
\end{figure*}

Fig.~\ref{fig:overview} summarizes the stationary branches and uniform-state stability boundaries. At $\beta=0.2$, the two nonuniform families join the uniform ordered state at the spatial thresholds $D_l$ and $D_u$ in Eq.~\eqref{eq:spatial-thresholds}. The shaded map records stability of the uniform ordered state to $m=\pm1$ perturbations and the exact incoherent threshold $D_{\rm inc}$. It does not exclude coexistence with nonuniform states, whose continuation and stability are examined in Sec.~\ref{sec:continuation}.

Fig.~\ref{fig:profiles-dynamics}(a,b) illustrates the spatial structure of representative states on these branches. Along the displayed noisy chimera branch, increasing noise reduces the contrast in local coherence and broadens the phase distributions in the strongly coherent region, as illustrated by the stationary phase samples in Fig.~\ref{fig:profiles-dynamics-supplement}(c)--(d).

Direct kinetic evolution connects these stationary profiles to relaxation across the stability diagram. All five runs in Fig.~\ref{fig:profiles-dynamics}(c) start from the same slightly perturbed noisy chimera at $D_0=0.001$, with only $D$ varied; panels (d)--(f) show three representative spatial evolutions. The initial condition and numerical protocol are specified in Appendix~\ref{sec:kinetic-details}.

At $D=0.001$, $0.005$, and $0.02$, spatial coherence contrast persists through $T=600$. At $D=0.05$, the profile becomes spatially uniform with $R(T)\simeq0.587479$, whereas at $D=0.12$ it approaches incoherence with $R(T)\simeq1.19\times10^{-6}$. The former retains collective order, while the latter approaches $f_{\rm inc}=1/(2\pi)$, consistent with the distinct spatial and incoherent thresholds $D_u$ and $D_{\rm inc}$. Across the three lower-noise runs, the final global coherence increases while the spatial contrast decreases, so $R$ alone does not identify a chimera. These finite-time observations complement the branch and spectral calculations; they do not independently locate phase boundaries or establish complete relaxation of every run.

\section{Spatial bifurcations and critical relaxation}\label{sec:amplitude}
The spectrum locates the loss of stability but does not determine whether a small nonuniform state grows continuously from the uniform branch or turns toward its stable side. We now retain the leading nonlinear feedback, including the change in collective rotation frequency, and compare the resulting amplitude dynamics with independent stationary and time-dependent calculations.

\subsection{Critical modes and symmetry reduction}\label{sec:modes}
At either endpoint, the cosine and sine spatial copies of the critical mode define a complex amplitude $B$, distinct from the global oscillator phase $\psi$. The expansion uses the near-threshold ordering $D-D_c=O(|B|^2)$. We use the phase generator $(Tu)_n=\ii n u_n$ and normalize the critical eigenvector $q$ so that $S=|B|+O(|B|^3)$. Projection conventions are given in Appendix~\ref{sec:solvability}.

Writing $a=(a_n)_{n\ge1}$, the complex spatial amplitude $B$ is defined by
\begin{equation}\label{eq:Bdef}
 \begin{aligned}
 a(x,t)=e^{T\psi(t)}\bigl[&U(D)+q\Rea\{Be^{2\pi\ii x}\}\\
 &+O(|B|^2)\bigr]
 \end{aligned}
\end{equation}
in the frame rotating at $\Omega_h(D)$. The phase $\arg B$ describes the position of the spatial pattern. It is distinct from the global oscillator phase $\psi$. Under the action $a(x)\mapsto a(x+\chi/(2\pi))$, $B\mapsto e^{\ii\chi}B$; spatial reflection sends $B$ to $\conj B$. With $\mu=D-D_c$, these symmetries imply
\begin{align}
 \dot B={}&s\mu B+c_3|B|^2B\nonumber\\
 &+O(\mu^2|B|+|\mu||B|^3+|B|^5),\label{eq:normalform}\\
 \dot\psi={}&\nu|B|^2+O(|\mu||B|^2+|B|^4).\label{eq:phase}
\end{align}
The coefficients $s,c_3,\nu$ are real. Thus the laboratory collective rotation frequency is $\Omega_h(D)+\dot\psi$. Reflection forbids a spatial drift term in this local one-mode normal form. Relative-equilibrium reduction and phase reconstruction follow the framework of Krupa \cite{Krupa}; oscillator Fokker--Planck amplitude expansions were developed by Crawford \cite{Crawford}. We apply these methods to the nonlocal kinetic hierarchy to determine the coefficients below.

For the reflection-symmetric family, $|b|=|B|+O(|B|^3)$, so the coefficients below use the cosine-field amplitude measured in the simulations.

\subsection{Nonlinear coefficients and the two endpoint types}\label{sec:endpoints}
The quadratic nonlinearity in Eq.~\eqref{eq:hierarchy} generates a uniform correction $h_0$ and a second spatial harmonic $h_2$ at order $|B|^2$. A bordered solve fixes the global-phase gauge and determines the frequency shift $\nu$ together with $h_0$. Projection onto the left nullvector $p$ of $L_1$, normalized by $p^{\mathsf T}q=1$ in the real-vector representation, then gives
\begin{equation}\label{eq:cubic}
 \boxed{\begin{aligned}
 c_3=p^{\mathsf T}\bigl[&2Q_{10}(q,h_0)\\
                      &+Q_{12}(q,h_2)-\nu Tq\bigr].
 \end{aligned}}
\end{equation}
Here $Q$ is half the Hessian of the moment vector field, with subscripts indicating the spatial harmonics of its arguments. The three contributions are feedback from the uniform correction, feedback from the second spatial harmonic, and phase-frequency reconstruction. Appendix~\ref{sec:solvability} supplies the explicit bilinear form, the bordered systems, and the derivative along the uniform branch that determines $s$.

The projected coefficients distinguish the lower and upper endpoint types. Numerical coefficients in the reduced equations below are rounded to six significant digits; Table~\ref{tab:coefficients} in Appendix~\ref{sec:solvability} lists their higher-precision values.

The phase-frequency term is essential: at the upper endpoint, omitting it gives $c_3\simeq+0.236568$, whereas the full result is $c_3\simeq-0.932544$. Fig.~\ref{fig:mechanism} in Appendix~\ref{sec:solvability} resolves this sign change and the convergence of independent dynamic estimates.

For $r=|B|$, the nonzero equilibrium satisfies $r^2=-s\mu/c_3$. At the lower endpoint,
\begin{equation}
 \begin{aligned}
 \dot B={}&7.34552(D-D_l)B\\
          &+0.181384|B|^2B+\cdots,\\
 |b|^2\simeq{}&40.4970(D_l-D).
 \end{aligned}
\end{equation}
The small nonuniform branch lies on the stable-uniform side $D<D_l$ and has radial eigenvalue $-2s\mu>0$. It is subcritical and radially unstable. A stable large-amplitude chimera or the width of a hysteresis loop cannot be deduced from this cubic local equation alone.

At the upper endpoint, it is clearer to use $\epsilon=D_u-D$:
\begin{equation}
 \begin{aligned}
 \dot B={}&2.63612\epsilon B\\
          &-0.932544|B|^2B+\cdots,\\
 |b|\simeq{}&1.68131\sqrt{D_u-D}.
 \end{aligned}
\end{equation}
The small nonuniform branch is radially stable, modulo its global-phase and translation symmetries. This is a supercritical spatial bifurcation when $\epsilon$ increases. Increasing noise reverses it and continuously restores spatial uniformity. The predicted leading radial relaxation rate is $-5.27224(D_u-D)$, sufficiently close to the endpoint.

\subsection{Spatial symmetry restoration and a Landau-type description}\label{sec:landau}
The continuum equations retain spatial translation and reflection symmetry for all $D$; the transition changes the symmetry of the state. A nonuniform density on the chimera-connected branch selects a pattern position, whereas the uniform ordered density is invariant under every spatial translation. Thus the upper endpoint restores spatial translation symmetry of the mean-field density. The corresponding order parameter is the complex modulation amplitude $B$, with $S=|B|+O(|B|^3)$, rather than the global coherence $R$. Since $R$ remains nonzero, this transition is distinct from restoration of phase-distribution uniformity at $D_{\rm inc}$. Spatial contrasts in coherence do not imply a local symmetry under independent transformations at each position.

At cubic order, let $r=|B|$, $a_u\simeq2.63612$, and $g_u\simeq0.932544$. The radial dynamics can be written as
\begin{equation}\label{eq:landau-potential}
 \begin{aligned}
 V(r)&=-\frac{a_u\epsilon}{2}r^2+\frac{g_u}{4}r^4,\\
 \dot r&=-\frac{dV}{dr},\qquad \epsilon=D_u-D.
 \end{aligned}
\end{equation}
Within this near-threshold approximation, for $D<D_u$ the minima of $V(|B|)$ form a circle in the complex-$B$ plane, with radius $\sqrt{a_u\epsilon/g_u}$ and arbitrary pattern position. For $D\ge D_u$, the minimum is at $B=0$. This provides a nonequilibrium Landau-type amplitude description of the continuous spatial transition, within the general framework of near-threshold pattern formation \cite{CrossHohenberg1993}.

The function $V$ is a local effective potential for the reduced modulation dynamics, not an identified thermodynamic free energy of the oscillator system. The collective phase still evolves according to Eq.~\eqref{eq:phase}, and stationary phase currents need not vanish. This construction therefore neither establishes detailed balance nor assigns a thermodynamic temperature to $D$. Its square-root scaling describes the local mean-field bifurcation; finite-amplitude folds require higher-order terms or full continuation.

\begin{figure*}[t]
\centering
  \begingroup
  \begin{lrbox}{\resultfigurebox}%
    \input{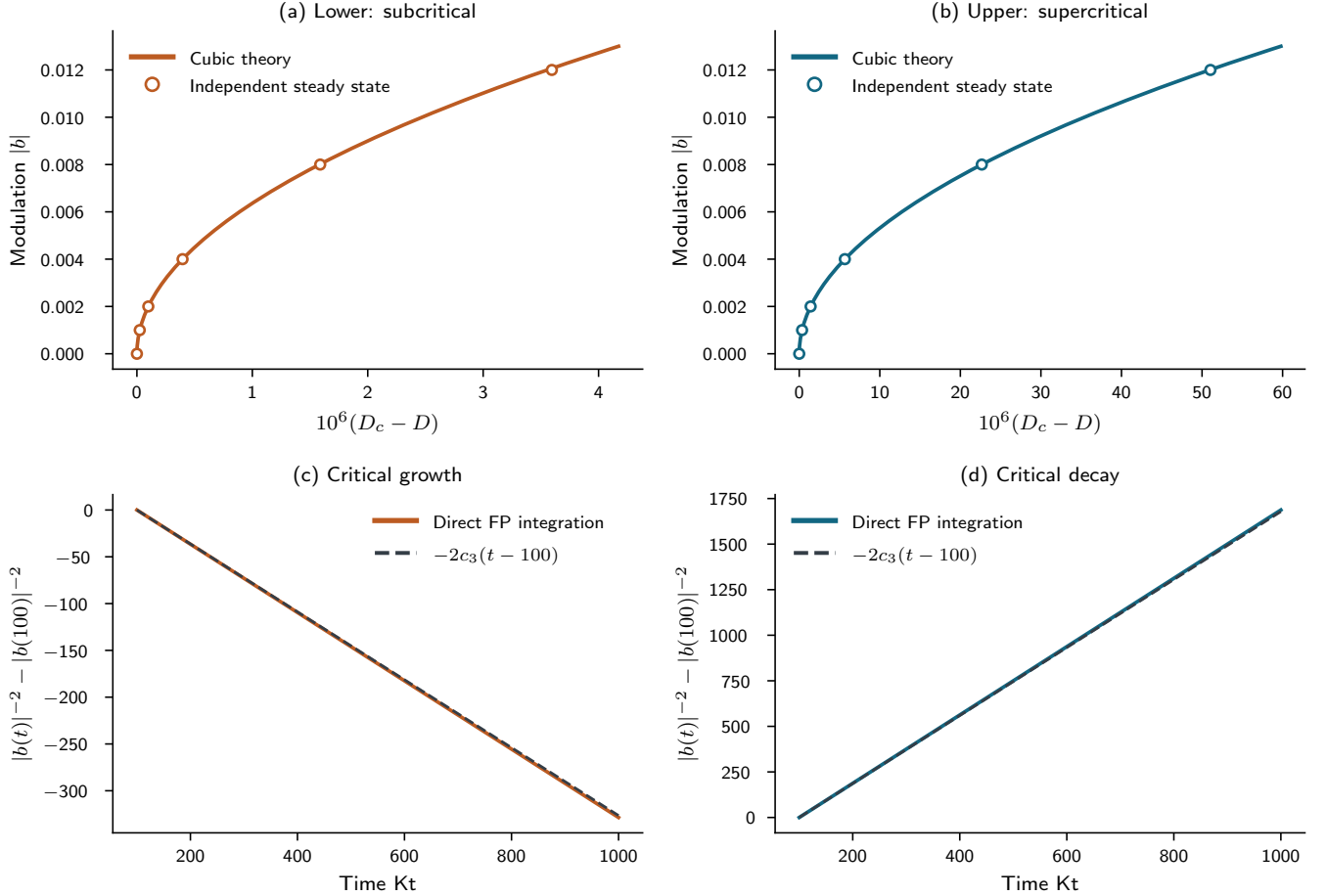}%
  \end{lrbox}%
  \resizebox{\textwidth}{!}{\usebox{\resultfigurebox}}%
  \endgroup

\caption{Independent validation of the local amplitude equation for $A=0.95$, $\beta=0.2$. (a),(b) Field amplitude near the lower and upper endpoints: solid curves give the cubic prediction and open circles give the independent stationary self-consistency calculation at $M=96$. The lower branch is subcritical and radially unstable; the upper branch is supercritical and radially stable near its endpoint. (c),(d) Direct Fokker--Planck integration at $D=D_c$, starting with $|b(0)|=0.01$, compared with $|b(t)|^{-2}-|b(100)|^{-2}=-2c_3(t-100)$. The time integrations use $M=64$, $N_x=32$, and $\Delta t=0.02$. The opposite slopes correspond to nonlinear growth and decay, respectively.}
\label{fig:validation}
\end{figure*}

\subsection{Independent validation and near-critical relaxation}\label{sec:near-upper-dynamics}
Independent stationary continuation recovers the predicted amplitude and frequency shifts, with quadratic branch slopes agreeing with $-c_3/s$ to better than $10^{-9}$ relative error. Direct critical-noise evolution gives nonlinear growth at the lower endpoint and decay at the upper endpoint (Fig.~\ref{fig:validation}). The polynomial fits and spectral-gap checks are collected in Appendix~\ref{sec:solvability}.

To test the transition itself, we integrate the full hierarchy at $D=D_u+\delta D$, with $\delta D=\{-5,-1,0,1,5\}\times10^{-4}$, through $T=10^4$. All five runs start from the same stationary density with $S(0)=0.02$. Below $D_u$, modulation approaches the independently computed nonuniform branch; above $D_u$, it decays while $R$ remains nonzero. The destination is uniform partial synchronization. Full numerical protocols and terminal values are given in Appendix~\ref{sec:near-critical-checks}.

At $D=D_u$, Eq.~\eqref{eq:normalform} predicts, to leading order,
\begin{equation}\label{eq:critical-relaxation}
 S(t)^{-2}-S(t_0)^{-2}
 \simeq -2c_3(t-t_0)\simeq1.86509(t-t_0).
\end{equation}
The independently measured slope over $2000\le t\le10^4$ is $1.87321$, differing by $0.44\%$.
This supports the predicted shifted algebraic relaxation toward zero; the nonzero terminal amplitude at criticality is a finite-time residue, not a stationary nonuniform state.
For $\delta D>0$, late relaxation instead follows the leading uniform-state eigenvalue, $S\propto e^{\Lambda_1(D)t}$.
The fitted late-time rates agree with the independently computed eigenvalues within $1\%$ for the two positive offsets. The slower near-critical relaxation explains why terminal snapshots alone can misidentify the endpoint.

Refinement checks cover three representative cases through $t=2000$, and density reconstructions remain positive at the sampled checkpoints. These symmetric trajectories test the amplitude dynamics; stability to general perturbations is assessed separately by the spectra.

\begin{figure*}[t]
\centering
  \begingroup
  \begin{lrbox}{\resultfigurebox}%
    \input{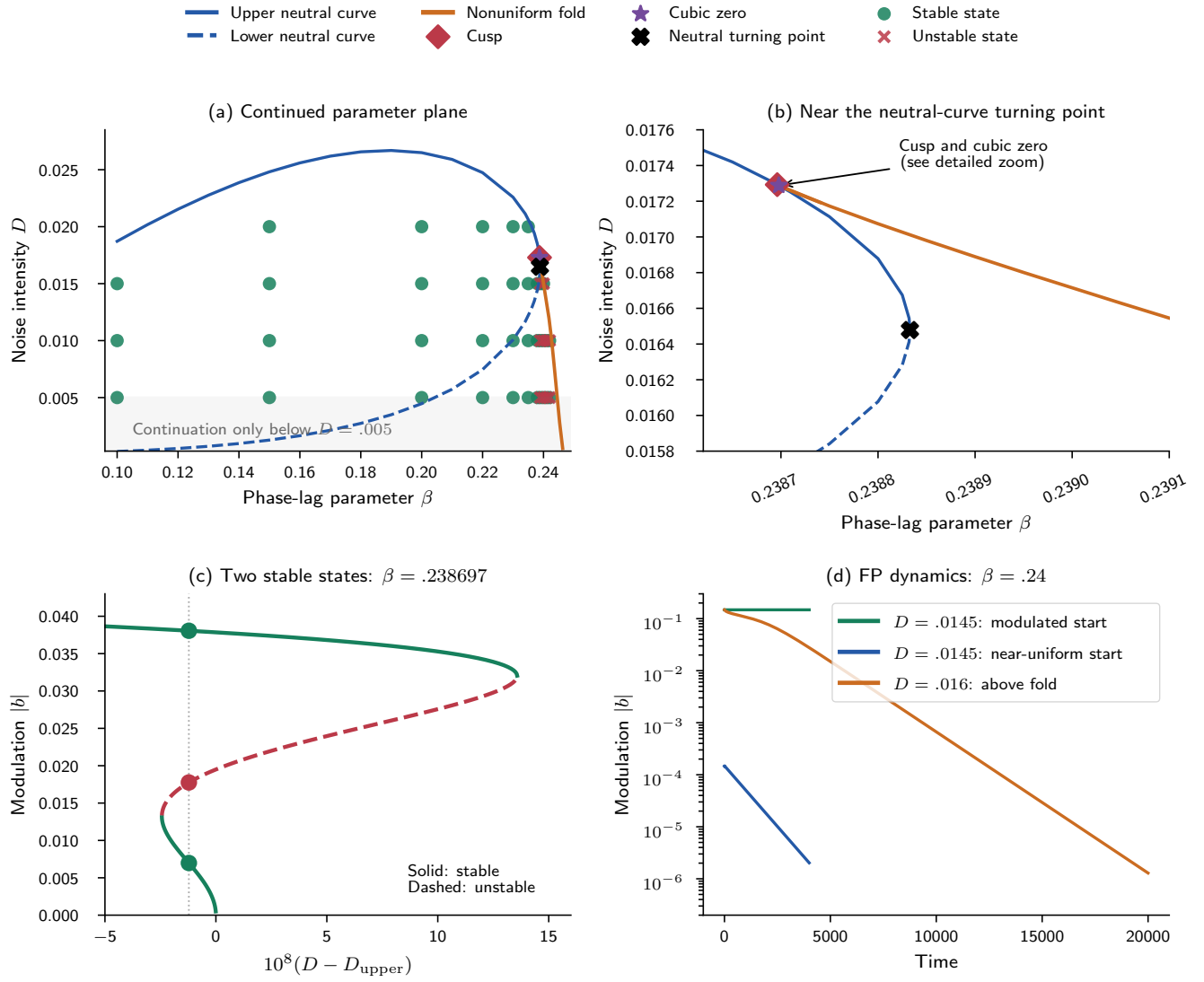}%
  \end{lrbox}%
  \resizebox{\textwidth}{!}{\usebox{\resultfigurebox}}%
  \endgroup

\caption{Two-parameter continuation at $A=0.95$. (a) Uniform-state neutral curves, the nonuniform saddle-node curve, and nonuniform states with numerically checked stability. The shaded low-noise region is not exhaustively audited for stability. (b) The neighborhood of the nonuniform cusp, cubic-zero point, and turning point of the uniform neutral curve; Fig.~\ref{fig:cuspzoom} resolves the separation of the first two. (c) At $\beta=0.238697$, two stable nonuniform branches are separated by an unstable branch; solid and dashed curves indicate the inferred stable and unstable branch segments. Full spectra at the marked states exclude the neutral global-phase and translation modes. Here $D_{\rm upper}$ is the upper uniform-state neutral threshold at this $\beta$. Markers show the states listed in Table~\ref{tab:radial}. (d) At $\beta=0.24$, nonuniform and near-uniform initial conditions approach nonuniform and uniform states, respectively, at $D=0.0145$, whereas the nonuniform initial state loses its modulation at $D=0.016$, above the saddle node.}
\label{fig:phaseplane}
\end{figure*}

\begin{table*}[tb]
\caption{Distinct parameter structures near the change in the spatial bifurcation type at $A=0.95$. Extra digits specify the reference coordinates for the small-offset tests in Appendix~\ref{sec:cusp-details}.}
\label{tab:specialpoints}
\begin{ruledtabular}
\begin{tabular}{lrr}
Structure & $\beta$ & $D$\\
\hline
Nonuniform cusp & $0.2386961858$ & $0.01729281154$\\
Cubic-zero spatial bifurcation & $0.2386977819$ & $0.01728818141$\\
Turning point of the uniform neutral curve & $0.2388328894$ & $0.01647845151$\\
\end{tabular}
\end{ruledtabular}
\end{table*}

\clearpage

\section{Two-parameter continuation and finite-amplitude loss}\label{sec:continuation}
We now vary $\beta=\pi/2-\alpha$ at fixed $A=0.95$. We continue the single-headed chimera family as reflection-symmetric nonuniform rotating densities, including its weakly modulated continuation near uniform states. Fixed-amplitude self-consistency and pseudo-arclength continuation resolve folds; stability calculations include both even and odd spatial perturbations. The main continuation covers $\beta=0.10$ to approximately $0.2463$ with $D\ge0.0005$. This parameter range includes both uniform-state spatial bifurcations and finite-amplitude folds of the continued family.

The continued boundaries, selected stability checks, and representative dynamics are summarized in Fig.~\ref{fig:phaseplane}.

\subsection{Finite-amplitude loss and coexistence}
The uniform-state neutral curve does not bound all nonuniform states. At $\beta=0.24$, a finite-amplitude saddle node occurs at
\begin{equation}
 \begin{aligned}
 D_{\rm SN}&\simeq0.0151044,\\
 |b|_{\rm SN}&\simeq0.129085.
 \end{aligned}
\end{equation}
The spatially uniform ordered state is stable there. A stable nonuniform state and a uniform state coexist on the audited segment immediately below this fold, as confirmed by spectra and separate initial conditions. Increasing $D$ through the fold destroys the continued nonuniform state at finite amplitude. This provides a discontinuous route to spatial uniformity, in contrast to the continuous upper endpoint at $\beta=0.2$.

\subsection{Cusp geometry and the cubic-zero bifurcation}
Three nearby parameter structures must be distinguished (Table~\ref{tab:specialpoints}).
The cusp has $|b|\simeq0.02456$.

The finite-amplitude cusp is distinct from both the cubic-zero point and the turning point of the uniform neutral curve. Appendix~\ref{sec:cusp-details} resolves their geometry and the associated small growth rates.

At the cubic-zero point, the next term in the radial normal form is destabilizing:
\begin{equation}\label{eq:quintic-summary}
 \begin{aligned}
 \dot B&=\lambda B+c_3|B|^2B+c_5|B|^4B+\cdots,\\
 c_5&\simeq+0.162512.
 \end{aligned}
\end{equation}
Here $\lambda$ is the critical linear growth rate. The coefficient is extracted at $c_3=0$ from the static branch and the known dynamic linear scale, rather than a general fifth-order dynamic reduction. Its positive sign requires higher-order terms or the full branch to determine finite-amplitude saturation. In the effective-potential representation, this sign gives a negative sixth-order term $-c_5r^6/6$, so the cubic-zero point is not identified with a conventional Landau tricritical point stabilized by a positive sixth-order term. Appendix~\ref{sec:cusp-details} gives the expansion and independent fold-scaling checks.

The cusp creates two stable nonuniform amplitudes separated by an unstable one, as resolved at $\beta=0.238697$ in Fig.~\ref{fig:phaseplane}(c). Their radial rates can be as small as $10^{-9}$, explaining why ordinary time sweeps cannot reliably resolve this coexistence. The eigenvalues and convergence checks are listed in Appendix~\ref{sec:cusp-details}. No Hopf bifurcation or stable breathing state was confirmed in the examined states.

\section{Discussion and conclusions}\label{sec:conclusions}
Independent phase noise changes spatial organization and global phase coherence at distinct thresholds. At $\beta=0.2$, the upper spatial transition occurs at $D_u\simeq0.02649537$, well below $D_{\rm inc}\simeq0.0993347$, and leads to uniform partial synchronization. This is a restoration of spatial translation symmetry, with a nonequilibrium Landau-type description in terms of the modulation amplitude and its local effective potential. The lower threshold marks loss of spatial stability of the uniform ordered branch rather than a minimum noise required for chimera existence. The spatial spectrum, phase-aware cubic reduction, and direct kinetic relaxation consistently identify a subcritical lower endpoint and a supercritical upper endpoint. Including the collective-frequency correction is essential to recover the upper-endpoint cubic sign and its continuous loss of modulation.

Continuation in phase lag extends this local mechanism to finite-amplitude saddle-node loss and a narrow region of nonuniform multistability. The cusp, cubic-zero point, and turning point of the uniform neutral curve are distinct structures. At the cubic-zero point, the positive quintic coefficient requires the full branch or higher-order terms to determine saturation.

These conclusions concern the continued single-headed family at $A=0.95$ and the stated parameter and observation ranges. The numerical spectra and refinements support the local reduction without establishing an infinite-dimensional center-manifold theorem or excluding other attractors; no breathing state was confirmed in the examined set. Within this scope, the results connect stationary branch geometry, spatial symmetry restoration, and critical relaxation in a common kinetic description.

\section*{acknowledgement}
The author acknowledges support from GP-Spin at Tohoku University.

\bibliography{main}

\clearpage
\appendix
\makeatletter
\@addtoreset{figure}{section}
\makeatother
\renewcommand{\thefigure}{\Alph{section}\arabic{figure}}
\renewcommand{\theHfigure}{appendix.\Alph{section}.\arabic{figure}}
\section{Kinetic construction, spectra, and numerical methods}\label{sec:kinetic-details}\label{sec:numerical-benchmarks}
\subsection{Kinetic equation and density reconstruction}
For a smooth periodic test function $h(\theta)$, It\^o's formula gives
\begin{equation}
 \frac{d}{dt}\int_0^{2\pi}h f\,d\theta
 =\int_0^{2\pi}(v h'+D h'')f\,d\theta.
\end{equation}
Integration by parts gives Eq.~\eqref{eq:fp}.

The numerical truncation of Eq.~\eqref{eq:hierarchy} retains $1\le n\le M$ and sets $a_{M+1}=0$. The associated density reconstruction is
\begin{equation}\label{eq:density-reconstruction}
 \begin{aligned}
 f_\Omega^{(M)}(x,\vartheta,t)
 =\frac{1}{2\pi}\biggl[1+2\Rea\sum_{n=1}^M
 a_n(x,t)e^{-\ii n\vartheta}\biggr].
 \end{aligned}
\end{equation}
The zeroth moment ensures normalization, but a finite Fourier truncation does not automatically preserve nonnegativity. We therefore check the reconstructed density as well as moment convergence.

\subsection{Stationary continued-fraction response}
For the stationary field in Eq.~\eqref{eq:stationary-field}, we use the cosine-kernel self-consistency construction \cite{Abrams2006}, with the noisy local response computed from Eq.~\eqref{eq:hierarchy}. For $D>0$, let $H(x)=e^{-\ii\alpha}Z_*(x)$ and introduce the continued-fraction ratios $\eta_n=a_n/a_{n-1}$. The truncated stationary hierarchy yields
\begin{equation}\label{eq:continued-fraction}
 \begin{aligned}
 \eta_{M+1}&=0,\\
 \eta_n(x)&=\frac{H(x)}{2[nD+\ii\Omega]+\conj{H(x)}\eta_{n+1}(x)},\\
 a_n(x)&=\prod_{k=1}^{n}\eta_k(x).
 \end{aligned}
\end{equation}
The backward recurrence avoids exponentials that become ill-conditioned at small noise. It is closed by Eq.~\eqref{eq:self-consistency}.

These are four real conditions for $(c,\Rea b,\Ima b,\Omega)$: the imaginary part of $\langle a_1\rangle_x$ must vanish. The sine average vanishes by reflection symmetry. Reflection is imposed to construct this stationary family, while time evolution and stability calculations retain both even and odd spatial perturbations. The noiseless locked and drifting conditional distributions are handled separately when initializing $D=0$ particle simulations.

\subsection{Uniform and nonuniform spectral operators}
The phase-incoherent density is $f_{\rm inc}=1/(2\pi)$. Linearizing the first phase moment in a spatial Fourier mode $m$ in the laboratory frame gives
\begin{equation}\label{eq:incoherent-spectrum}
 \lambda_m=-D+\tfrac12 e^{-\ii\alpha}g_m,
\end{equation}
where the cosine-kernel multipliers are
\begin{equation}\label{eq:kernel-multipliers}
 g_0=1,\qquad g_{\pm1}=A/2,\qquad g_{|m|\ge2}=0.
\end{equation}
Higher phase harmonics $n\ge2$ have decay rates $-Dn^2$. Because $0\le A\le1$ and $\sin\beta>0$ in the parameter range considered, the leading incoherent growth rate is $-D+\sin\beta/2$, giving Eq.~\eqref{eq:incoherent-threshold}.

Using the kernel multipliers in Eq.~\eqref{eq:kernel-multipliers}, the linear block for a perturbation in a cosine or sine spatial harmonic is
\begin{align}
 (L_m u)_n={}&(-Dn^2-\ii n\Omega_h)u_n\nonumber\\
 &+\frac n2\bigl(H_hu_{n-1}-\conj H_hu_{n+1}\bigr)\nonumber\\
 &+\frac n2g_m e^{-\ii\alpha}u_1U_{n-1}\nonumber\\
 &-\frac n2g_m e^{\ii\alpha}\conj{u_1}U_{n+1}.\label{eq:linear}
\end{align}
Here $H_h=e^{-\ii\alpha}c_h$ and $u_0=0$, whereas $U_0=1$. The phase generator is $(Tu)_n=\ii n u_n$, and $L_0TU=0$. This neutral global-phase mode belongs to $m=0$ and is separate from the spatial instability.

A nonuniform stationary density couples spatial harmonics, so a uniform-state mode map alone cannot determine its stability. At fixed $\Omega$, let $h(x,\vartheta)$ be a real density perturbation satisfying
$\int_0^{2\pi}h(x,\vartheta)\,d\vartheta=0$ at every $x$. The linearized kinetic operator is
\begin{equation}\label{eq:full-linearization}
 \mathcal L_*h=-\partial_\vartheta
 \bigl(v_{\Omega,*}h+f_*\delta v[h]\bigr)
 +D\partial_\vartheta^2h,
\end{equation}
where
\begin{equation}\label{eq:velocity-variation}
 \begin{aligned}
 \delta v[h]&=\Ima\!\left[e^{-\ii(\vartheta+\alpha)}\delta Z[h](x)\right],\\
 \delta Z[h](x)&=\int_0^1G(x-y)\,\delta z[h](y)\,dy,\\
 \delta z[h](y)&=\int_0^{2\pi}e^{\ii\vartheta}h(y,\vartheta)\,d\vartheta.
 \end{aligned}
\end{equation}
In moment variables this operator is real-linear: both $\delta Z$ and $\conj{\delta Z}$ must be varied. Keeping only a complex-linear derivative would miss part of the feedback.

Global phase invariance gives the neutral tangent $-\partial_\vartheta f_*$. A spatially nonuniform family also has the translation tangent $-\partial_x f_*$. These symmetry modes must be identified when interpreting eigenvalues near zero. For a spatially uniform density, $\partial_x f_*=0$ and supplies no additional neutral vector; at a spatial bifurcation the cosine and sine critical modes instead enter independently. Stability in the main text refers to decay of nonsymmetry perturbations within the numerical resolution, with the mass constraint enforced. Dense spectra at coarse resolution are combined with targeted eigenvalue windows at finer resolution. The refined windows check the relevant weakly damped or unstable modes, rather than constituting an exhaustive computation of the infinite-dimensional spectrum. Section~\ref{sec:modes} uses the critical spatial modes to separate the global phase from the modulation amplitude.

\subsection{Time integration and benchmark checks}
We retain $M$ positive phase Fourier moments, set $a_{M+1}=0$, and evaluate spatial averages on $N_x$ equidistant periodic grid points. Eq.~\eqref{eq:three-spatial-moments} evaluates the cosine-kernel convolution through three spatial averages. Stationary solutions use the backward recurrence~\eqref{eq:continued-fraction} and field closure~\eqref{eq:self-consistency}; the continuation constraints are specified with the corresponding branch calculations. Time-dependent solutions use fourth-order exponential time differencing (ETDRK4) \cite{Kassam2005}, with the diagonal phase-diffusion term $-Dn^2a_n$ integrated exactly. This removes the explicit stability restriction from the stiff diffusion term, but still requires time-step convergence and density checks.

We specify $(N_x,M,\Delta t,T)$ with each experiment because the required phase resolution changes substantially with noise strength. The checks below give low-noise density, time-step, and incoherent-spectrum benchmarks. Appendix~\ref{sec:solvability} gives the separate local-coefficient and near-critical checks.

For the initial five relaxation benchmarks at $\beta=0.2$, we used
$(N_x,M,\Delta t,T)=(256,192,0.01,600)$; the subsequent calculations use the resolutions stated with their results. At $D=0.001$, refining a stationary nonuniform solution from $(N_x,M)=(256,128)$ to $(512,192)$ changes its four field parameters by at most $3.7\times10^{-14}$. The latter reconstructed density has minimum $-1.1\times10^{-14}$ and normalization error below $9\times10^{-16}$, consistent with roundoff. For this same low-noise state, $M=64$ produces visible negative density values of order $10^{-2}$, illustrating why convergence must be checked for each state rather than inferred from a single truncation.

In a separate perturbed-state evolution at $D=0.001$ with $(N_x,M)=(128,128)$, the maximum spatial difference in $a_1(x,T)$ between $\Delta t=0.01$ and $0.005$ is $5.3\times10^{-13}$ at $T=100$. An independent small-perturbation test about incoherence reproduces the exact amplification factors $e^{\lambda_m T}$ for $m=0,1,2$ to relative error below $3\times10^{-11}$ at $D=0.06$, $T=10$, and $\Delta t=0.01$. These are convergence checks for the specified benchmarks, not uniform error bounds for the complete phase diagram. The branch, normal-form, and particle comparisons in the main text provide separate checks for their respective calculations.

\subsection{Additional profiles and common initial state}
\begin{figure*}[t]
\centering
  \begingroup
  \begin{lrbox}{\resultfigurebox}%
    \input{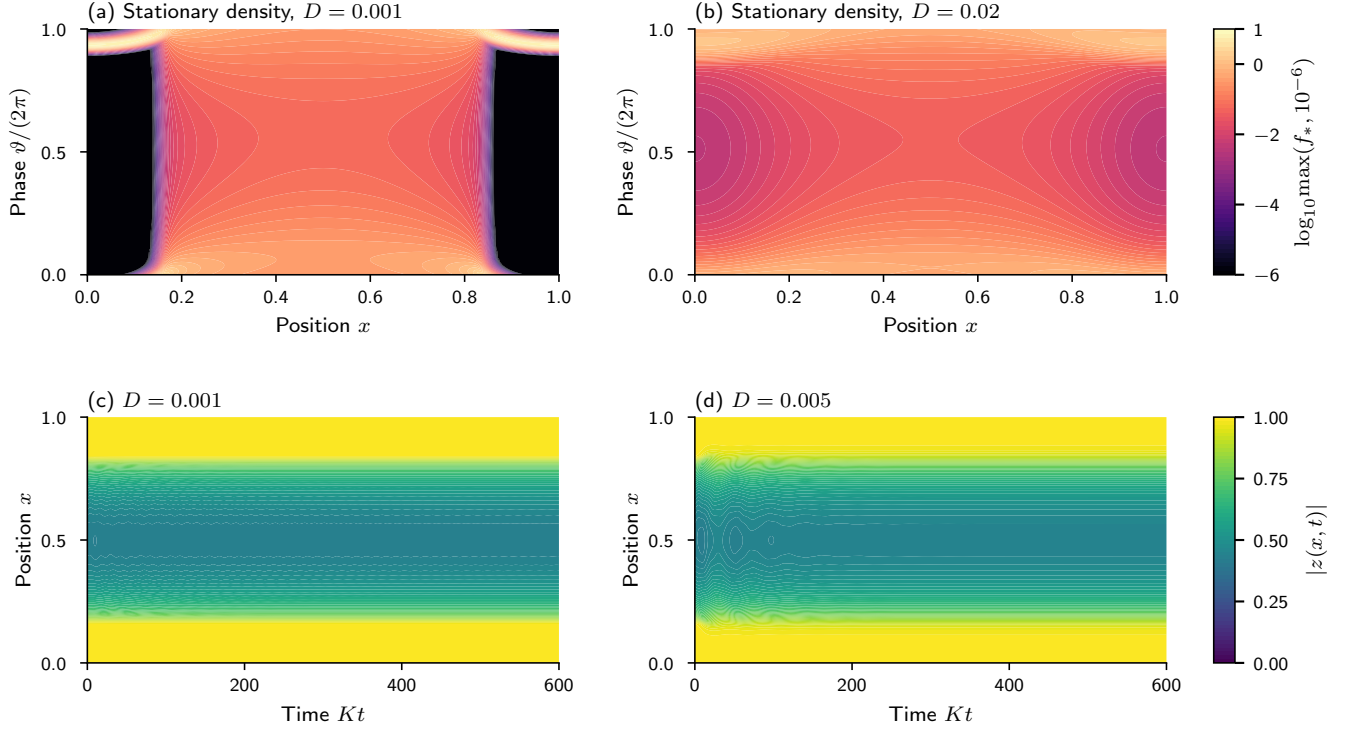}%
  \end{lrbox}%
  \resizebox{\textwidth}{!}{\usebox{\resultfigurebox}}%
  \endgroup

\caption{Additional stationary and dynamical profiles at $A=0.95$, $\beta=0.2$, and $K=1$, complementing Fig.~\ref{fig:profiles-dynamics}. (a,b) Rotating-frame stationary densities on branch I at $D=0.001$ and $0.02$, with $\vartheta=\theta-\Omega t$, normalized with respect to $d\vartheta$. Both use $\log_{10}[\max(f_*,10^{-6})]$; the floor is for display only. (c,d) Local coherence $|z(x,t)|$ at $D=0.001$ and $0.005$, on the same 0--1 scale as the main-text evolution maps. All runs use the common initial condition in Eq.~\eqref{eq:fp-initial-condition}. Calculations use $N_x=256$ and $M=192$; densities are evaluated at 720 phase points. Across this figure and Fig.~\ref{fig:profiles-dynamics}, all five trajectories use ETDRK4 with $\Delta t=0.01$ up to $T=600$, with every saved snapshot at unit time intervals shown in the maps.}
\label{fig:profiles-dynamics-supplement}
\end{figure*}

Fig.~\ref{fig:profiles-dynamics-supplement}(a,b) displays the rotating stationary phase densities underlying the spatial observables in Fig.~\ref{fig:profiles-dynamics}(a,b). Their broadening with noise complements the reduction in spatial coherence contrast.

Together, Figs.~\ref{fig:profiles-dynamics}(d)--(f) and \ref{fig:profiles-dynamics-supplement}(c,d) show the five relaxation benchmarks from a common initial condition; Fig.~\ref{fig:profiles-dynamics}(c) gives their global order parameter. We take the stationary noisy chimera on branch I at $D_0=0.001$ and apply the same small spatial phase perturbation to all phase moments:
\begin{equation}\label{eq:fp-initial-condition}
 \begin{aligned}
 a_n(x,0)&=a_{n,*}(x;D_0)e^{\ii n\delta\vartheta(x)},\\
 \delta\vartheta(x)&=0.005\bigl[\cos(6\pi x)+\sin(4\pi x)\bigr].
 \end{aligned}
\end{equation}
Only the noise intensity $D$ is changed between integrations; the initial density, kernel, phase lag, and numerical resolution are held fixed. These are deterministic evolutions of the nonlinear Fokker--Planck equation for an oscillator population subject to independent phase noise.

\section{Amplitude solvability and independent validation}\label{sec:solvability}

\subsection{Projection conventions and bordered equations}
At either critical $D_c$, $L_1$ has a simple zero eigenvalue in the internal real phase-vector space. The cosine and sine spatial copies produce two critical real modes. Choose right and left vectors $q,p$ such that
\begin{equation}
 \begin{gathered}
 L_1q=0,\qquad p^{\mathsf T}L_1=0,\\
 p^{\mathsf T}q=1,\qquad |g_1q_1|=1,\qquad \Rea q_1>0.
 \end{gathered}
\end{equation}
The transpose and pairing refer to the real-vector representation. The global phase zero mode is removed separately, by requiring the spatial mean of $a_1$ to be real.

For the reflection-symmetric representative $B=r\in\mathbb R$, the nonlocal field has $Z=c+b\cos(2\pi x)$ and $b=g_1q_1r+O(r^3)$ after fixing the global phase. Consequently $|b|=|r|+O(|r|^3)$. The cubic coefficient is normalized to this cosine-field amplitude. Using the one-sided Fourier amplitude $|\widehat Z_1|=|b|/2$ multiplies $c_3$ and $\nu$ by four.

Write $F(U+u)=L u+Q(u,u)$ at fixed $D,\Omega_h$, so $Q$ is half the Hessian. For internal vectors associated with spatial multipliers $g_j,g_\ell$,
\begin{align}
 [Q_{j\ell}(u,v)]_n={}&\frac n4e^{-\ii\alpha}g_j u_1v_{n-1}\nonumber\\
 &+\frac n4e^{-\ii\alpha}g_\ell v_1u_{n-1}\nonumber\\
 &-\frac n4e^{\ii\alpha}g_j\conj{u_1}v_{n+1}\nonumber\\
 &-\frac n4e^{\ii\alpha}g_\ell\conj{v_1}u_{n+1}.\label{eq:Q}
\end{align}
Perturbation zeroth phase moments are zero. Let
\begin{equation}
 \begin{aligned}
 u={}&rq\cos(2\pi x)\\
 &+r^2\bigl(h_0+h_2\cos(4\pi x)\bigr)+O(r^3),\\
 \dot\psi={}&\nu r^2+O(r^4).
 \end{aligned}
\end{equation}
Using $\cos^2(2\pi x)=[1+\cos(4\pi x)]/2$, the second-order equations are
\begin{align}
 L_0h_0-\nu TU&=-\tfrac12 Q_{11}(q,q),\label{eq:h0}\\
 L_2h_2&=-\tfrac12 Q_{11}(q,q).\label{eq:h2}
\end{align}
With the gauge condition $\Ima(h_{0,1})=0$, Eq.~\eqref{eq:h0} is a bordered system, not an inversion of the singular $L_0$. For a left phase nullvector $p_0$, it gives
\begin{equation}
 \nu=\frac{p_0^{\mathsf T}Q_{11}(q,q)}{2p_0^{\mathsf T}TU}.
\end{equation}
At third order, $2\cos(2\pi x)\cos(4\pi x)=\cos(2\pi x)+\cos(6\pi x)$ determines the feedback into the critical harmonic.
The resulting projection is Eq.~\eqref{eq:cubic} in the main text.
There is no explicit cubic nonlinearity in \eqref{eq:hierarchy}; the cubic amplitude term is generated by slaved modes and phase reconstruction.

With $\mu=O(r^2)$ and $\dot r=O(r^3)$, time derivatives of the second-order slaved terms enter at order $r^4$ and do not change Eq.~\eqref{eq:cubic}. The spatial reflection symmetry restores Eq.~\eqref{eq:normalform} from the cosine calculation without discarding the sine critical mode.

The linear slope must be differentiated along the uniform ordered branch:
\begin{equation}
 s=p^{\mathsf T}\frac{dL_1}{dD}q.
\end{equation}
In particular, derivatives of $U(D)$ and $\Omega_h(D)$ must be included. They follow from a second bordered solve:
\begin{equation}\label{eq:UD}
 \begin{pmatrix}L_0&-TU\\ \gamma^{\mathsf T}&0\end{pmatrix}
 \begin{pmatrix}U_D\\ \Omega_{h,D}\end{pmatrix}
 =\begin{pmatrix}(n^2U_n)_n\\0\end{pmatrix}.
\end{equation}
Here $\gamma^{\mathsf T}u=\Ima u_1$. All complex entries in this expression are represented by real and imaginary parts in the actual solve.

\subsection{Coefficient values and phase-frequency feedback}
At $M=96$ the projection calculation gives the values in Table~\ref{tab:coefficients}; the displayed precision is supported by checks through $M=192$.
\begin{table}[tb]
\caption{Amplitude-reduction coefficients at $\beta=0.2$ and $A=0.95$. The amplitude is normalized to the cosine component of the nonlocal field.}
\label{tab:coefficients}
\begin{ruledtabular}
\begin{tabular}{lrr}
Quantity & Lower endpoint & Upper endpoint\\
\hline
$D_c$ & $0.00446591038$ & $0.02649536903$\\
$c_h(D_c)$ & $0.9644669190$ & $0.7654209155$\\
$\Omega_h(D_c)$ & $-0.9357758351$ & $-0.7599351287$\\
$s$ & $+7.34551751$ & $-2.63612173$\\
$c_3$ & $+0.181384107$ & $-0.932543939$\\
$\nu$ & $+2.549518136$ & $+3.213387966$\\
$\Rea h_{0,1}$ & $-2.946818184$ & $-5.487943782$\\
\end{tabular}
\end{ruledtabular}
\end{table}

The cubic decomposition is particularly informative:
\begin{align}
 c_3^{\rm lower}&=10.719394389-0.583137990\nonumber\\
                &\quad-9.954872293\nonumber\\
                &\simeq+0.181384107,\\
 c_3^{\rm upper}&=0.976269405-0.739701869\nonumber\\
                &\quad-1.169111475\nonumber\\
                &\simeq-0.932543939.
\end{align}
At the upper endpoint, omitting phase-frequency feedback would give a positive coefficient $+0.236567536$ and the wrong local transition type.

\begin{figure*}[t]
\centering
  \begingroup
  \begin{lrbox}{\resultfigurebox}%
    \input{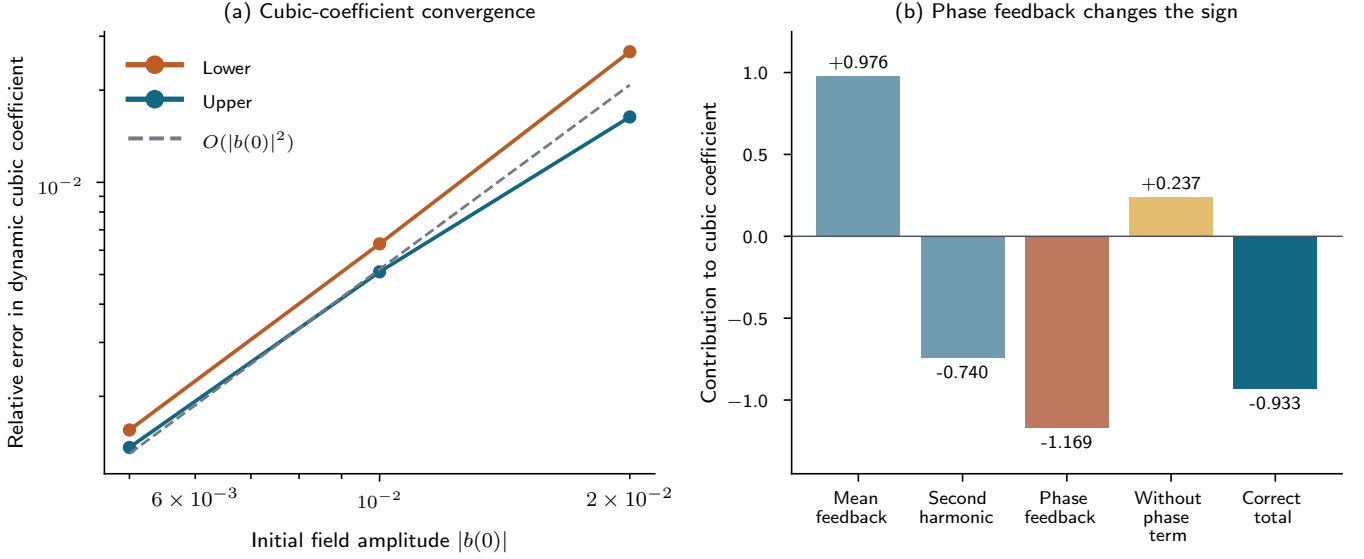}%
  \end{lrbox}%
  \resizebox{\textwidth}{!}{\usebox{\resultfigurebox}}%
  \endgroup

\caption{Convergence and mechanism of the cubic amplitude coefficient at $A=0.95$, $\beta=0.2$. (a) Relative error of the coefficient inferred from direct Fokker--Planck evolution at each critical noise intensity, fitted over $100\le t\le1000$, versus initial field amplitude $|b(0)|$. The dashed line indicates an $O(|b(0)|^2)$ trend. Calculations use $(M,N_x,\Delta t)=(96,64,0.01)$ for $|b(0)|=0.005,0.02$ and $(64,32,0.02)$ for $|b(0)|=0.01$. (b) The three contributions to Eq.~\eqref{eq:cubic} at the upper endpoint, their sum without phase reconstruction, and the full coefficient. Including the phase-frequency contribution changes the sign of $c_3$.}
\label{fig:mechanism}
\end{figure*}

Fig.~\ref{fig:mechanism} shows the convergence of the dynamic cubic estimates and the contributions responsible for this sign change.

\subsection{Stationary fits and spectral separation}\label{sec:checks}
An independent continuation fixes $q_b=|b|$ and solves for $(c,\arg b,\Omega,D)$, using a local continued-fraction stationary response rather than the adjoint projection. With 45-digit arithmetic it gives
\begin{align}
 D-D_l&=-0.0246931692178\,q_b^2\nonumber\\
      &\quad-1.90329557714\,q_b^4+\cdots,\\
 D-D_u&=-0.353756022806\,q_b^2\nonumber\\
      &\quad-4.97062801138\,q_b^4+\cdots.
\end{align}
The quadratic slopes agree with $-c_3/s$ from the projection to better than $10^{-9}$ relative error. The independently measured frequency shifts satisfy
\begin{equation}
 \Omega-\Omega_h(D)=\nu q_b^2+O(q_b^4)
\end{equation}
with the coefficients in Table~\ref{tab:coefficients}. The static $q_b^4$ coefficient is not itself the dynamic quintic Landau coefficient.

The phase Fourier truncation was increased through $M=48,64,96,128,192$. At $M=96$, after excluding the appropriate zero eigenvalues, the largest real parts of the $m=0,1,|m|\ge2$ blocks are respectively about
\begin{equation}
 \begin{aligned}
 \text{lower: }&(-0.16942,-0.28763,-0.16925),\\
 \text{upper: }&(-0.14258,-0.26019,-0.18096)
 \end{aligned}
\end{equation}
for the lower and upper endpoints. Thus the numerical spectrum supports the assumed local spectral separation. Direct time integration at $D=D_c$ independently finds growth near the lower endpoint and decay near the upper endpoint; the observed coefficients approach the projected ones as the initial modulation decreases. Data, discretization checks, and plots accompany this manuscript.

\subsection{Near-critical integration protocol and finite-time data}\label{sec:near-critical-checks}
To test the dynamical consequences of the upper-endpoint reduction, we integrate the full hierarchy~\eqref{eq:hierarchy} at $A=0.95$, $\beta=0.2$, and $D=D_u+\delta D$, where
$\delta D=\{-5,-1,0,1,5\}\times10^{-4}$.
All five runs start from the same positive, reflection-symmetric stationary density on branch I at $D_0=0.0263530677739$.
The common initial modulation is $S(0)=0.02$, with $S$ defined in Eq.~\eqref{eq:observables}. For this reflection-symmetric family $S=|b|$, and $S=|B|$ to leading order near the endpoint.
ETDRK4 integration uses $(N_x,M,\Delta t)=(64,64,0.02)$ up to $T=10^4$, retaining both cosine and sine field feedback.
The evolution is computed in a common frame with $\Omega=\Omega_h(D_u)$; both $S$ and $R$ are invariant under this frame choice.

Table~\ref{tab:near-upper-dynamics} shows the finite-time outcomes.
Below $D_u$, the evolution approaches independently computed stationary nonuniform densities.
At $\delta D=-5\times10^{-4}$, the terminal amplitude agrees with the stationary value $0.0372289378556$ to relative difference $5.7\times10^{-12}$.
At $\delta D=-10^{-4}$, it remains $0.071\%$ above the stationary value $0.0167798471307$.
Above $D_u$, spatial modulation decays while global coherence remains nonzero: the destination is uniform partial synchronization, not phase incoherence.

\begin{table}[t]
\caption{Finite-time direct Fokker--Planck results near $D_u=0.0264953690303$. All entries are evaluated at $T=10^4$ and are not automatically stationary values.}
\label{tab:near-upper-dynamics}
\begin{ruledtabular}
\begin{tabular}{rrr}
$\delta D$ & $S(T)$ & $R(T)$\\
\hline
$-5\times10^{-4}$ & $0.03722894$ & $0.76169042$\\
$-10^{-4}$ & $0.01679174$ & $0.76466182$\\
$0$ & $0.00685746$ & $0.76516251$\\
$10^{-4}$ & $0.00091057$ & $0.76462490$\\
$5\times10^{-4}$ & $2.66\times10^{-8}$ & $0.76146915$\\
\end{tabular}
\end{ruledtabular}
\end{table}

Fits over $6000\le t\le10^4$ give rates $-2.67042\times10^{-4}$ and $-1.34024692\times10^{-3}$ for $\delta D=10^{-4}$ and $5\times10^{-4}$, respectively, compared with the independently computed eigenvalues $-2.64494\times10^{-4}$ and $-1.34024687\times10^{-3}$.
The slower near-critical relaxation explains why terminal snapshots alone can misidentify the endpoint.

A joint refinement to $(N_x,M,\Delta t)=(128,96,0.01)$ at $\delta D=-5\times10^{-4},0,5\times10^{-4}$ changes $a_1(x,t)$ by at most $2.9\times10^{-11}$ over $0\le t\le2000$.
Density reconstructions at checkpoints separated by $1000$ time units remain positive, with minimum exceeding $0.014$, and normalization errors remain below $5\times10^{-16}$.
The refinement comparison covers these three runs through $t=2000$, not the entire $10^4$ interval.
These symmetric trajectories test the local amplitude dynamics; stability to general perturbations is assessed separately by the spectral calculations.

\section{Resolution of the cusp and cubic-zero point}\label{sec:cusp-details}
\begin{figure*}[t]
\centering
  \begingroup
  \begin{lrbox}{\resultfigurebox}%
    \input{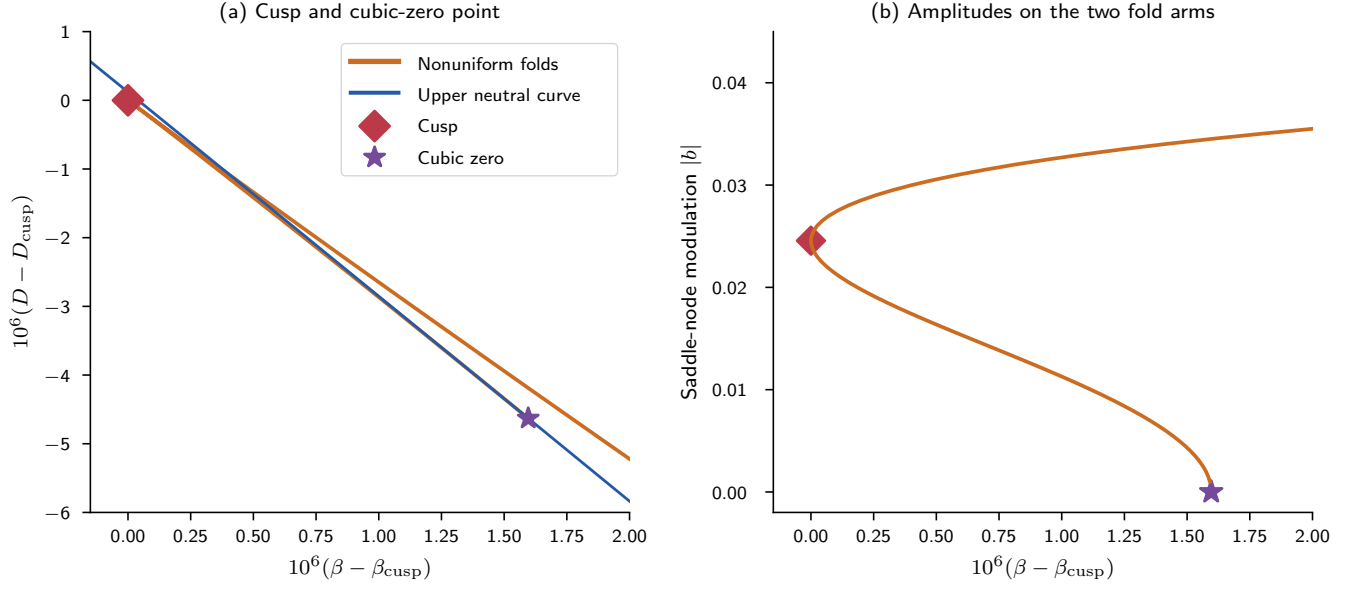}%
  \end{lrbox}%
  \resizebox{\textwidth}{!}{\usebox{\resultfigurebox}}%
  \endgroup

\caption{Resolved cusp geometry at $A=0.95$. (a) Nonuniform fold loci and the upper uniform-state neutral curve in coordinates centered on the nonuniform cusp. The red diamond marks $(\beta_{\rm cusp},D_{\rm cusp})$ and the purple star marks the distinct cubic-zero point (Table~\ref{tab:specialpoints}). (b) The corresponding saddle-node amplitudes. The two finite-amplitude folds meet at the cusp, while the small-amplitude fold terminates at zero amplitude at the cubic-zero point. Axis offsets are multiplied by $10^6$ to display their separation.}
\label{fig:cuspzoom}
\end{figure*}

Fig.~\ref{fig:cuspzoom} resolves the two nearby folds and their distinction from the cubic-zero bifurcation.
 At the neutral-curve turning point, $\lambda=\partial_D\lambda=0$ but $\partial_\beta\lambda\simeq-0.503735\ne0$; this is not an additional double-zero dynamical eigenvalue.

Denote the cubic-zero coordinates in Table~\ref{tab:specialpoints} by $(\beta_*,D_*)$. At this point, $s=-0.163042086944$ and $c_3=0$. A 45-digit independent self-consistency expansion gives
\begin{equation}
 D(|b|)=D_*+0.996747561923|b|^4+O(|b|^6).
\end{equation}
Since the cubic term vanishes, the critical quintic coefficient is
\begin{equation}
 \begin{aligned}
 \dot B&=\lambda B+c_3|B|^2B+c_5|B|^4B+\cdots,\\
 c_5&=-s(0.996747561923)\\
    &=+0.162511802652.
 \end{aligned}
\end{equation}
The positive sign is independently converged in truncation and fitting window. It excludes an interpretation based on a stabilizing negative quintic alone. Higher-order terms and the full branch determine finite-amplitude saturation. This extraction of $c_5$ uses the static branch and the known dynamic linear time scale at $c_3=0$; it is not a complete fifth-order dynamic center-manifold derivation away from that point.

An independent 40-digit continuation further checks the small-fold predictions $S_{\rm fold}^2=-c_3/(2c_5)$ and $\lambda_{\rm fold}=c_3^2/(4c_5)$, where $S_{\rm fold}=|b|_{\rm fold}$. As $\beta-\beta_*$ decreases in magnitude from $-10^{-7}$ to $-10^{-8}$, the actual-to-predicted amplitude ratio changes from $1.007774$ to $1.000759$, and the noise-displacement ratio from $1.008729$ to $1.000860$.

The narrow cusp structure creates two stable nonuniform amplitudes separated by an unstable one. For example, at $\beta=0.238697$, $D=0.0172904996292$, the three amplitudes and leading radial eigenvalues are given in Table~\ref{tab:radial}.
\begin{table}[tb]
\caption{Coexisting nonuniform states at $\beta=0.238697$ and $D=0.0172904996292$. Neutral global-phase and translation modes are excluded.}
\label{tab:radial}
\begin{ruledtabular}
\begin{tabular}{rrl}
$|b|$ & $\lambda_{\rm rad}$ & Stability\\
\hline
$0.00697866345$ & $-3.25894\times10^{-9}$ & stable\\
$0.01775646409$ & $+1.72192\times10^{-8}$ & unstable\\
$0.03805927119$ & $-4.30199\times10^{-7}$ & stable\\
\end{tabular}
\end{ruledtabular}
\end{table}
Global phase and translation modes are removed using the symmetry tangents before these eigenvalues are identified. Three full-matrix resolutions agree; the other modes remain stable. The extremely small rates explain why ordinary finite-time sweeps cannot resolve this structure reliably. No Hopf bifurcation or stable breathing state was confirmed in the audited states; this statement does not exclude such states elsewhere.

\end{document}